\documentclass[12pt]{article}
\usepackage{dina4p}
\usepackage{amssymb}
\usepackage{graphicx}

\newcommand{\BE}{\begin{equation}}
\newcommand{\EE}{  \end{equation}}
\newcommand{\BEA}{\begin{eqnarray}}
\newcommand{\EEA}{  \end{eqnarray}}

\newcommand{\PD}[1]{\partial_{#1}}
\newcommand{\pd}[2]{\frac{\partial #1}{\partial #2}}

\newcommand{\BF}[1]{{\mathbf{#1}}}
\newcommand{\MC}[1]{{\mathcal{#1}}}

\newcommand{\R}{ {\mathbb R}} % the reals
\newcommand{\Z}{ {\mathbb Z}} % the integers
\newcommand{\T}{ {\mathbb T}} % the torus

\newcommand{\ra}{\rightarrow }

\newcommand{\BV}{\left(\begin{array}{c}}     % := ``begin{vector}''
\newcommand{\EV}{\end{array}\right)}
\newcommand{\BM}[1]{\left(\begin{array}{#1}} % := ``begin{matrix}''
\newcommand{\EM}{\end{array}\right)}

\newcommand{\DOF}{degrees-of-freedom}

\newcommand{\POT}{\Phi}
\newcommand{\TRA}{^{\rm T}}
\newcommand{\PRJ}{{\mathcal P}}
\newcommand{\LAT}{{\mathcal M}}
\newcommand{\MODE}{f^\pm}
\newcommand{\LMODE}{\hat{f}^\pm}
\newcommand{\SMODE}{\hat{g}^\pm}
\newcommand{\SETA}{\eta^\pm}

\begin{document}
\title{A New Model for the Collective Beam-Beam Interaction}
\author{J A Ellison \\
        UNM, Albuquerque, \\ NM 87131, USA \\
        {\tt ellison@math.unm.edu}
   \and A V Sobol \\
        Tech-X Corporation, \\ Boulder, CO 80303, USA \\
        {\tt sobol@txcorp.com}
   \and M Vogt \\
        DESY--MPY, \\ 22607 Hamburg, FRG\\
        {\tt vogtm@mail.desy.de}
}
\date{}

\maketitle

\begin{abstract} 
        The Collective Beam-Beam interaction is studied in the framework of 
maps with a ``kick-lattice'' model in the 4-D phase space of the transverse motion.  
        A novel approach to the classical method of averaging is used to derive 
an approximate map which is equivalent to a flow within the averaging approximation.
        The flow equation is a continuous-time Vlasov equation which we call
the averaged Vlasov equation, the new model of this paper. 
        The power of this approach is evidenced by the fact that the averaged
Vlasov equation has exact equilibria and the associated linearized equations have uncoupled azimuthal
Fourier modes.
The equation for the Fourier modes leads to a Fredholm integral equation of the third kind and 
the setting is ready-made for the development of a weakly nonlinear
theory to study the coupling of the $\pi$ and $\sigma$ modes.
The $\pi$ and $\sigma$ eigenmodes are calculated from the third kind integral equation. 
These results are compared with the kick-lattice model using our weighted macroparticle tracking code
 and a newly developed, density tracking, parallel, Perron-Frobenius code.
 
\end{abstract}

{\small
% 02.30.Rz = integral equations
% 29.20.Dh = storage rings
% 29.27.Bd = beam-dyn, collective eff, instabilities
% 45.20.Jj = Lagrangian & Hamiltonian mech
% 52.59.-f = intense particle beams and radiation sources
% 52.65.-y = Fokker-Planck & Vlasov equation
Submitted to \emph{New Journal of Physics}}

\section{               Introduction          }\label{sc:intro}
In this paper we introduce a new model for the collective beam-beam
 interaction for hadron beams in 4D transverse phase space (2 \DOF).
This both generalizes and simplifies the work of \cite{MS,CR00,YO00,AL00} on the 
 collective beam-beam interaction in high energy colliders.
In addition, it extends the preliminary 1 degree-of-freedom collective case in \cite{EV01}.
Our model is based on the classical method of averaging generalized to maps and collective forces.
We do not distribute the beam-beam force around the ring as is usually done.
The technique we introduce should be of general interest for studies of 
Vlasov systems with a localized perturbative collective force.

In Section \ref{sc:kicklatt}, we discuss our basic kick-lattice model for the evolution of the 4D phase 
 space densities of the two beams.
In Section \ref{sc:error}, we briefly review the basic averaging theory which is generalized in this paper.
Previously, this averaging theory was applied to the weak-strong beam-beam in one and 
 two degrees-of-freedom \cite{DEV02,ElDuSaSeSoVo}.
The equations of the kick-lattice model will be transformed to a standard form for the method 
 of averaging in Section \ref{sc:AVE} and the general ``averaged Vlasov equation'' (AVE) will be derived
 (See equation (\ref{eq:AVE})).
We then introduce the special case we treat in this paper, namely the case where the tunes of the two beams 
 are identical and are non-resonant.
In this case the AVE has the property that any function of the action only is an equilibrium solution.
In Section \ref{sc:LAVE}, we linearize about these equilibria and discuss the linearized equations and the 
 associated third kind integral equation.
In addition, we compare our integral equation with the analogous integral equations which arise in
 the standard plasma problem and in the beam dynamics problems concerning the longitudinal dynamics with
 wake fields for a coasting beam and a bunched beam.  
In Section \ref{sc:num} we present numerical results for the $\pi$ and $\sigma$ mode eigen-problems
 for an axially symmetric Gaussian equilibrium
 and compare these results with simulations on the exact model of Section \ref{sc:kicklatt}.
In Section 7 we give a summary and point to future work.
An appendix is included which gives a first principles calculation of the beam-beam force.    

\section{The Kick-Lattice Model in 4D Phase Space}\label{sc:kicklatt}

        To describe the evolution equations,  we refer to the bunches as 
``unstarred'' and ``starred'', and for every quantity $X$ describing the 
unstarred bunch, the quantity $X^*$ describes the starred bunch. 
        The evolution equations are symmetric: the equation for the starred bunch is
obtained from the unstarred bunch by interchanging starred and unstarred quantities, 
so we mostly state only the equation for the unstarred bunch.

  We consider two counter-rotating particle bunches, which collide head on 
at a single interaction point (IP). 
The electromagnetic interaction at the IP is determined up to a proportionality factor by the dimensionless
 ``potential'' $\POT$, which satisfies the Poisson equation $-\Delta \POT =2\pi \rho^*$. 
        Here $\rho^*(x,y)$ is the spatial density (normalized to one),
and the  potential $\POT[\rho^*]:\R^2\ra\R$ is given by
\BE
\POT[\rho^*](x,y) =\int_{\R^2} \,\,\, G(x-x',y-y')
 \rho^* (x',y')\,dx'dy',
\label{eq:POT}
\EE
where  $G(x,y)=-\ln(\sqrt{x^2+y^2}/\sigma)=-\ln(\sqrt{x^2+y^2})+\ln(\sigma)$ is the Green's function.
In the following we will omit the scaling factor $\sigma$ which is in principle needed for
 dimensional correctness but which can be chosen completely arbitrarily since it does not contribute to the
 beam-beam kick.

        Letting $n$ refer to the state of the system just before the IP, 
particles in the unstarred bunch change  their phase-space position $u=(x,y,p_x,p_y)\TRA$
according to the map
\BE
%\fl
 u_{n+1}  =  \LAT \left( u_n + \zeta 
      \left(
        0, 0, \pd{}{x}\POT[\rho^*_n](\PRJ u_n),\pd{}{y} \POT[\rho^*_n](\PRJ u_n) 
      \right)\TRA
    \right).
\label{eq:KR}
\EE
The associated phase space density $\psi_n$ evolves via $\psi_{n+1}(u_{n+1}) = 
\psi_n(u_n)$, or
\BE
%\fl
 \psi_n(u)      =  \psi_{n+1}\left(\LAT \left( u + \zeta 
      \left(
        0, 0, \pd{}{x}\POT[\rho^*_n](\PRJ u), \pd{}{y} \POT[\rho^*_n](\PRJ u) 
       \right)\TRA
    \right)\right),
\label{eq:EVSold}
\EE
 which is easily inverted to give
\BE
%\fl
 \psi_{n+1}(u)       =
 \psi_{n}\left( u - \zeta 
      \left(
        0, 0, \pd{}{x} \POT[\rho^*_n](\PRJ \LAT^{-1}u), \pd{}{y} \POT[\rho^*_n](\PRJ \LAT^{-1}u) 
      \right)\TRA
    \right).
\label{eq:EVS}   % Exact Vlasov System
\EE
Here $\LAT$ is a stable linear symplectic map representing  the linear lattice,
 $\zeta$ is the beam-beam factor,
$ \PRJ=$ {\scriptsize $
   \BM{cccc}
      1 & 0 & 0 & 0 \\
      0 & 1 & 0 & 0
   \EM$ } projects phase space on configuration space,     
and the spatial and phase-space densities are related by
\BE
    \rho^*(x,y)=\int_{\R^2}  \psi^*(x,y,p_x,p_y) \,dp_x dp_y.
\EE
 The beam-beam factor, which is derived in the appendix, is
$\zeta=\frac{1+\beta\beta^*}{\beta+\beta^*}\frac{2N^*}{\beta\gamma}r_p$, where  
 the absolute value of $r_p=\frac{qq^*}{4\pi\epsilon_0mc^2}$ is the classical particle radius
 (as long as only elementary particles or ions of the same charge state are involved),
 $N$ is the number of particles, 
 $q$ is the particle charge,
 $\gamma$ is the Lorentz factor associated with $\beta$, and 
 $m$ is the particle mass.
For all modern colliders, i.e.\ in the limit $\beta$,$\beta^*$ $\to1$, $\zeta$ can be approximated 
 by  $\zeta\approx\frac{2N^*}{\gamma} r_p$.
The evolution law for the starred beam is obtained by replacing $\LAT$ by 
 $\LAT^*$,$\psi^*$ by $\psi$ and  $\zeta$ by $\zeta^*$ where starred and 
 unstarred are interchanged in $\beta$, $\gamma$, $N$ and $m$.
 
Equation (\ref{eq:KR}) can be written more compactly as
\BE
        u_{n+1}  =  \LAT\Big(u_n + \zeta \MC{J}_4 \nabla_u 
         \POT[\rho^*_n](\PRJ u_n)\Big),  
        \label{simp}
\EE
where 
  ${\mathcal J}_{2k}=$ {\scriptsize$\BM{rc}0_k&I_k\\-I_k&0_k\EM$} is the unit symplectic matrix.
We note here that a map is said to be symplectic if the Jacobian, $M$, of the map satisfies 
 $M\TRA {\mathcal J} M={\mathcal J}$.
We have written the kick in a ``Hamiltonian form'' because eventually a transformed
  $\POT$ will be a Hamiltonian for a flow.

For simplicity, we take 
\BE
   \LAT=\BM{cccc}
       C_x         &  0           & \beta_x S_x  & 0           \\
       0           &  C_y         & 0            & \beta_y S_y \\
      -S_x/\beta_x &  0           & C_x          & 0           \\
       0           & -S_y/\beta_y & 0            & C_y         \\
     \EM,
\EE
where $C_i:=\cos(2\pi\nu_i)$, $S_i:=\sin(2\pi\nu_i)$ and where $\nu_i$ for $i=x,y$ are the tunes. 
We have assumed that the beta functions, $\beta_x$ and $\beta_y$,  have  minima at the IP. 
The distinction between the lattice $\beta$ and the relativistic $\beta$
 should be clear from context.

To relate $\zeta$ to the usual beam-beam parameter, we linearize the kick
 in (\ref{eq:KR}) about $(x,y)=(0,0)$ in the case where $\rho^*$ is mirror symmetric and
 invariant under $\frac{\pi}{2}$ rotations,
 i.e.\ when $\rho^* (x,y) = \rho^* (-x,y) = \rho^* (x,-y) = \rho^* (y,x)$.
Note that this is still a weaker constraint than full axial symmetry.
Because of these symmetries, $\POT_x(0)=\POT_y(0)=\POT_{xy}(0)=0$ and 
 $\POT_{xx}(0,0) = \POT_{yy}(0,0) = -\pi \rho^*(0,0)$ where the latter uses
 Poisson's equation. 
Thus the kick matrix becomes {\scriptsize $\BM{cc} I& 0 \\ kI&I\EM$} where
 $k=\pi \zeta \rho^*(0,0)$.  
The tune shift is $\xi_i=\Delta \nu_i = -\frac{1}{4\pi} \beta_i k+O(k^2)$ with $i=x,y$.
Thus the beam-beam parameter $\xi_i=-\frac{1}{4} \beta_i \zeta \rho^*(0,0)$.
For a round Gaussian, this gives the standard result.

\section{       Map Averaging and Error Bounds       }\label{sc:error}

Here we give an overview of the averaging formalism, which we generalize in
 this paper, and briefly discuss error bounds.
We consider the autonomous ``kick-rotate'' map in ${\mathbb R}^2$ 
\BE
u_{n+1} = e^{\MC{J}_2 2\pi \nu}\left( u_n + \epsilon \big(0,-\POT'(u_{1,n})\big)\TRA \right)
\label{eq:KRM}
\EE
with the small parameter $\epsilon$.
This is a model for the one degree of freedom weak-strong beam-beam interaction and was discussed
 in \cite{DEV02}.
The transformation 
\BE
    u=e^{\MC{J}_2 2\pi n\nu} v
    \label{slowCart}
\EE
 leads to the non-autonomous map
\BE
v_{n+1} = v_n + \epsilon {\mathcal J}_2 \nabla_v H(v_n,n\nu),
\label{eq:SFFA}
\EE
where $H(v,\theta)=V(v_1\cos(2\pi\theta) + v_2\sin(2\pi\theta))$.
This is in a standard form for the method of averaging in which the transformed 
dependent variable, $v$, is slowly varying.
If $\nu$ is irrational, then from  Weyl's equidistribution theorem 
\cite{Koerner}
 the average of $H(v,n\nu)$ over $n$ exists and is given by
 $\bar{H}(v) = \int^1_0 H(v,\theta)\,d\theta$.
It is therefore natural to ask, for what values of $\nu$ are solutions of
 (\ref{eq:SFFA}) approximated by solutions of the averaged map
\BE
w_{n+1} = w_n + \epsilon {\mathcal J}_2 \nabla_w \bar{H}(w_n).
\label{discWS}
\EE
Even though the maps in (\ref{eq:KRM}) and (\ref{eq:SFFA}) are symplectic, the averaged map is not.
However the averaged flow associated with (\ref{discWS}) and defined by
\BE
\dot{w} = \epsilon {\mathcal J}_2 \nabla_w \bar{H}(w)
\label{contWS}
\EE
is Hamiltonian, and it is easy to show that 
$|w_n-w(n)|=O(\epsilon)$ over $O(1/\epsilon)$ times.
  Since (\ref{contWS}) is autonomous, (\ref{discWS}) can be viewed as the Euler
method for numerically integrating (\ref{contWS}).
Approximating  equation (\ref{eq:SFFA}) with (\ref{discWS}) is considered in our previous work \cite{DEV02},
 where we introduce the concept of a {\it far-from-low-order-resonance } zone for $\nu$. 
  This zone is formed
by removing a finite number of intervals centered on low-order rationals,
therefore $\nu$ needs to satisfy only finitely many Diophantine conditions, and 
does {\it not } need to be irrational, which makes the formalism
much more useful in the applications. 
  The error bound $|v_n-w_n|=O(\epsilon)$ is obtained without the usual
 near-identity-transformation and is unchanged asymptotically if an $O(\epsilon^2)$
 term is added to equations (\ref{eq:KRM},\ref{eq:SFFA}).

        In \cite{ElDuSaSeSoVo}, we extend the \emph{formalism} of equations 
(\ref{eq:KRM}-\ref{contWS}) to the weak-strong beam-beam interaction in 2 \DOF.
        The equation of motion corresponding to (\ref{eq:KRM}) is just equation
(\ref{eq:KR}) with $\rho^*_n$ replaced by the spatial density  of the strong beam.
        We are working out the details of the averaging theorem in this more complicated 
case with two frequencies  \cite{DEV03}. Some ingredients of our approach can be found
in \cite{DumEl}.
We generalize this to the collective beam-beam interaction in the next section.

\section{       Map-Averaging for Vlasov Systems      }\label{sc:AVE}

We will begin by transforming (\ref{eq:KR}) using a representation of the solution
 to the unperturbed, $\zeta =0$, problem.
The new coordinates will be slowly varying if $\zeta$ is small.
As in the previous section we could proceed by 
 letting $u=\LAT^nq$ which gives 
\BE
q_{n+1}=q_n+\zeta \LAT^{-n}{\mathcal J}_4 \nabla_u \POT[\rho_n^*](\PRJ \LAT^n q_n).
\label{eq:13}
\EE
The averaged equation then becomes 
\BE
 w_{n+1}=w_{n}+\zeta{\mathcal J}_4\nabla_w\bar{F}[\rho^*](w_n),
\EE
 where $\bar{F}[\rho^*](w)$ denotes the $n$-average of 
 $\LAT^{-n}{\mathcal J}_4 \nabla_u\POT[\rho^*](\PRJ \LAT^n w)$.

The transformation to (\ref{eq:13}) turned out to be a major advance in Sobol's implementation of 
 the Perron-Frobenius (PF) method \cite{SobolThesis} (See \cite{WE2000} and \cite{VESW01} for a 
 discussion of the PF method).
In addition, (\ref{eq:13}) is well suited for an error analysis which is in progress \cite{SobolAve}.
However, an action-angle transformation may be better suited to understand 
 approximate equilibria and the associated linear analysis that we do here and that is how we will proceed.

The action-angle transformation 
 from $u=(x,y,p_x,p_y)\TRA$ to  slowly varying coordinates $v=(\Theta_x,\Theta_y,J_x,J_y)\TRA$ is
 given by
%\numparts
\BEA
x   & = & \sqrt{2J_x\phantom{/}\beta_x} \sin(2\pi n\nu_x+\Theta_x)  \\
p_x & = & \sqrt{2J_x         / \beta_x} \cos(2\pi n\nu_x+\Theta_x)  \\
y   & = & \sqrt{2J_y\phantom{/}\beta_y} \sin(2\pi n\nu_x+\Theta_y)  \\
p_y & = & \sqrt{2J_y         / \beta_y} \cos(2\pi n\nu_y+\Theta_y). 
\EEA
%\endnumparts
Note that for fixed $J$ and $\Theta$ these are solutions of the equations of motion with $\zeta=0$, 
 that is without the beam-beam force.

Equation (\ref{eq:KR}) becomes
\BE
v_{n+1}=v_n+\zeta {\mathcal J}_4 \nabla_v H[\Psi^*_n](v_n,n)
 + O(\zeta^2),
\label{eq:KRAA}
\EE
where 
\BEA
%\fl
  H[\Psi^*](v,n) & := & \int_{\T^2\times\R^2_+} \Psi^*(v')\,dv'        \nonumber \\
     &\times& G\Big(\sqrt{2\beta_x   J_x }\sin(2\pi n\nu_x  +\Theta_x )-
                    \sqrt{2\beta_x^* J'_x}\sin(2\pi n\nu_x^*+\Theta'_x), \nonumber \\
     & & \phantom{G\Big(}\sqrt{2\beta_y   J_y }\sin(2\pi n\nu_y  +\Theta_y )-
                    \sqrt{2\beta_y^* J'_y}\sin(2\pi n\nu_y^*+\Theta'_y) \Big).
\label{eq:POTAA}
\EEA
The integral in (\ref{eq:POTAA}) is taken  over $[0,2\pi]$ in the
 $\Theta$'s and over $[0,\infty)$ in the $J$'s.
Since the transformation is symplectic it is also volume preserving.
Thus the $(\Theta,J)$-density is given by $\Psi_n(v)=\psi_n(u)$, and its evolution law is 
\BE
 \Psi_n(v) \;\,\,\, =  \Psi_{n+1}\Big(v \,+\, \zeta {\mathcal J}_4 \nabla_v H[\Psi^*_n](v,n)
                                  + O(\zeta^2)\Big), 
\EE
or equivalently
\BE
 \Psi_{n+1}(v)  =  \phantom{_{+1}}
                     \Psi_n\Big(v \,-\, \zeta {\mathcal J}_4 \nabla_v H[\Psi^*_n](v,n)
                                  + O(\zeta^2)\Big). 
\label{eq:AvMap}
\EE
Clearly $v_n$ and $v^*_n$ are slowly varying for $\zeta$ and 
 $\zeta^*$ small, and it
 follows that the transformed densities $\Psi$ and $\Psi^*$ are slowly varying.
Thus (\ref{eq:KRAA}) is in a standard form for averaging and 
 we now follow the procedure laid out in the previous section.
The averaged map problem is obtained from (\ref{eq:KRAA}) by replacing 
 $H$ by the appropriate $n$-average $\bar{H}$ and dropping the 
 $O(\zeta^2)$ term.
 The associated averaged flow problem is autonomous and has the  
 Hamiltonian form 
\BE
	\dot{w} = \zeta {\mathcal J}_4 \nabla_w \bar{H}[\Psi^*](w).
\EE
 Thus the averaged Vlasov equations for $\Psi$ and $\Psi^*$ become
\BE
\left\{ 
       \begin{array}{ccc}
          \PD{t} \Psi + \zeta \{\Psi,\bar{H}[\Psi^*]\} & = & 0  \\
          \PD{t} \Psi^* + \zeta^* \{\Psi^*,\bar{H}^*[\Psi]\} & = & 0
       \end{array}    
\right. ,               
\label{eq:AVE}
\EE
 where 
$\{f,g\}=
	 \pd{f}{\theta_x}\pd{g}{J_x}
	+\pd{f}{\theta_y}\pd{g}{J_y}
	-\pd{f}{J_x}\pd{g}{\theta_x}
	-\pd{f}{J_y}\pd{g}{\theta_y}
$ is the Poisson bracket.
Note that $H^*[\Psi]$ is obtained from (\ref{eq:POTAA}) by interchanging
 the starred and unstarred parameters $\beta$ and $\nu$. 
System (\ref{eq:AVE}) is the new model referred to in the title.
Since $\zeta$ and $\zeta^*$ are small one immediate advantage of 
 (\ref{eq:AVE}) over (\ref{eq:KR}) is that the step size in a numerical
 integration of (\ref{eq:AVE}) can be $O(1/\rm{max}(\zeta,\zeta^*))$ which
 is much larger than one turn.

At this stage the problem is general with parameters
 ($\nu_x,\nu_y,\nu_x^*,\nu_y^*,\beta_x,\beta_y,\beta_x^*,\beta_y^*,
   \zeta,\zeta^*$) and the correct averaged Hamiltonian $\bar{H}$ depends
 on the relation between the four tunes.
Here we discuss the case $\nu_x=\nu_x^*$ and $\nu_y=\nu_y^*$ because (i)
 we wish to compare and contrast our results with \cite{YO00,AL00} 
 and (ii) it simplifies 
 the calculation of the average.
In this case $G(\cdots)$ in (\ref{eq:POTAA}) can be rewritten as 
 $G(D_x\sin(2\pi n\nu_x+\varphi_x),D_y\sin(2\pi n\nu_y+\varphi_y))$, 
 where 
%\numparts
 \BEA 
  D_x = D(\beta_x J_x,\beta_x^* J^{'}_x,\Theta_x-\Theta^{'}_x), && \nonumber \\
  D_y = D(\beta_y J_y,\beta_y^* J^{'}_y,\Theta_y-\Theta^{'}_y), && \label{eq:convD} \\
  D(r,s,t) = \sqrt{2r+2s-4\sqrt{rs}\cos t}, &&
 \EEA
%\endnumparts
  and the phases $\varphi_x$ and $\varphi_y$
 are easily determined from the trigonometry involved.
If in addition we consider the case where $\nu_x$ and $\nu_y$ are
 non-resonant (in the sense that $k_x\nu_x+k_y\nu_y=k_0\Rightarrow k_x=k_y=k_0=0$),
 then the averaging over $n\nu_x$ and $n\nu_y$ can be done
 separately and each average can be replaced by the associated integral.
Thus the averaged Hamiltonian becomes
\BE
\bar{H}[\Psi^*](v)= \int_{\T^2\times\R_+^2} 
  \bar{G}\Big(D_x,D_y\Big)\Psi^*(v')\,dv',
\label{eq:AVHAM}
\EE
where 
$ \bar{G}(D_x,D_y):= 
   1/(2\pi)^2 \int_{\T^2}G(D_x \sin t_x, D_y\sin t_y) \,dt_x\,dt_y$
and as before $v=(\Theta, J)$.
From our experience with the non-collective case \cite{DEV02,ElDuSaSeSoVo}, we expect this to be valid 
 for $\nu_x$, $\nu_y$ far from low-order resonances; work on the error
 estimates is in progress \cite{SobolAve}.
Note that $H = H^*$ if $\beta_x=\beta_x^*$ and 
 $\beta_y=\beta_y^*$. 

Because of the convolution structure of the $\Theta$ integral in (\ref{eq:AVHAM}) 
 (see (\ref{eq:convD})) any function
 $\Psi^*=\Psi^*_e(J)$ results in $\bar{H}$ being 
 independent of $\Theta$.
It follows that any pair of densities $\Psi_e$ and $\Psi_e^*$ that are independent of $\Theta$
 are an equilibrium pair for (\ref{eq:AVE}). 
Since 
 $
  \psi_n(x,y,p_x,p_y)=\Psi_n(\Theta,J) \approx\Psi_e(J)=
  \Psi_e\left(\frac{1}{2}\left(\frac{x^2}{\beta_x}+\beta_x p_x^2\right),
              \frac{1}{2}\left(\frac{y^2}{\beta_y}+\beta_y p_y^2\right)\right)$,
we see that (\ref{eq:KR}) and equivalently  (\ref{eq:KRAA}) have 
 quasi-equilibria given the averaging approximation.
In \cite{EV01}, we have verified this in the 2D phase space case. 
\begin{figure}[htb]
\centering
\includegraphics*[width=8.5cm]{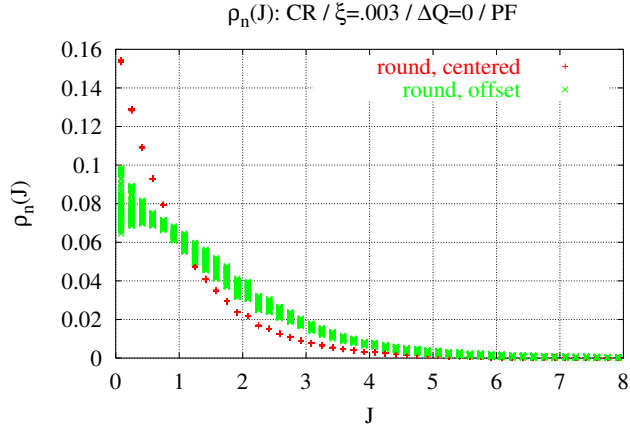}
\caption{The evolution of the action density in a 1 \DOF\ example.
         The red crosses are a quasi equilibrium while the green $\times$'s
         are a $\Theta$--dependent fluctuating density.
         The evolution is over $10^5$ turns.}  
\label{fig:RhoJ}
\end{figure}
Figure \ref{fig:RhoJ} shows the evolution of two action densities at a hundred out of
 more than $10^5$ turns under the exact map, the 2D analog of (\ref{eq:EVS}), using
 the PF method for tracking phase space densities.
The parameters are $\nu=\nu^*=\sqrt{5}-2$ and $\xi=3\times10^{-3}$.
The red crosses represent the evolution of a quasi-equilibrium, namely the centered Gaussian
 $\Psi_0(\Theta,J)=\Psi_e(J)=1/(2\pi\varepsilon)e^{-J/\varepsilon}$ (for $\varepsilon=1$),
 while the green $\times$'s stand for a Gaussian that was initially shifted by $1\sigma_x$,
 giving a $\Theta$-dependent density.
Since the red crosses for different discrete $J$ lie on top of each other,
 $\Psi_e$ hardly evolves over the $10^5$ turns which is consistent with the averaging result.
The $\Theta$-dependent density however, fluctuates as is indicated by the band of green $\times$'s.
Thus we have strong evidence for the existence of quasi-equilibria, however we believe that
 (\ref{eq:EVS}) does not admit exact equilibria. This is in contrast to the lepton case, see \cite{WaEl}.

\section{           The Linearized Equations          }\label{sc:LAVE}

To linearize about an equilibrium, we set $\Psi(v,t)=\Psi_e(J) + \Psi_1(v,t)$,
 and $\Psi^*(v,t)=\Psi^*_e(J) + \Psi^*_1(v,t)$.
Plugging into (\ref{eq:AVE}) and linearizing gives
\BE
\left\{
\begin{array}{l}
\PD{t}\Psi_1 + \zeta \{\Psi_1,\bar{H}[\Psi^*_e]\}
             + \zeta \{\Psi_e,\bar{H}[\Psi^*_1]\}=0,  \\
\PD{t}\Psi^*_1 + \zeta^* \{\Psi^*_1,\bar{H}^*[\Psi_e]\}
             + \zeta^* \{\Psi^*_e,\bar{H}^*[\Psi_1]\}=0.
\end{array}
\right.
\label{eq:LAVE}
\EE
Because of the convolution structure of $\bar{H}$ in (\ref{eq:AVHAM}),
 the Fourier modes of (\ref{eq:LAVE}) are uncoupled,
 but we do not pursue (\ref{eq:LAVE})
 in that generality here.
Instead we investigate solutions, which exist when
 $\Psi_e=\Psi_e^*$, $\zeta=\zeta^*$, and 
 $\bar{H}=\bar{H}^*$.
Letting $\MODE = \Psi_1 \pm \Psi^*_1$, we obtain
\BE
\PD{\tau}f^\pm  + \{\MODE,\bar{H}[\Psi_e]\}
            \pm \{\Psi_e,\bar{H}[\MODE]\} = 0,
\label{eq:Mode}
\EE
 where $\MODE$ are the so-called $\sigma$ (sum) and $\pi$ (difference) 
 modes respectively and we have scaled the $\zeta$ into the time
 by defining $\tau=\zeta t$.
We remind the reader that the densities $\Psi_1$ and $\Psi_1^*$ in (\ref{eq:LAVE}) are coupled,
 while $f^+$ and $f^-$ in (\ref{eq:Mode}) are not.
Moreover,  the $\sigma$ and $\pi$ modes can be intuitively interpreted as in-phase and 180$^\circ$ out-of-phase
 perturbations respectively to the starred and unstarred beam.
Note that $t$ in (\ref{eq:LAVE}) is dimensionless (same as $n$ in (\ref{eq:AvMap}))
 whereas $\tau$ in (\ref{eq:Mode}) has the dimension of a length.
Note also that the beam current, which is a factor of $\zeta$,  has been scaled out of the problem.
Thus there can be no threshold for instability in our problem.

Unfolding the Poisson brackets we obtain
\BE
 \PD{\tau}\MODE +\Omega(J)\cdot\nabla_\Theta\MODE \mp\nabla_J\Psi_e(J)\cdot\nabla_\Theta\bar{H}[\MODE]=0,
 \label{eq:LAVEtt} % LAVE in tau and Theta
\EE
 where
\BE
 \Omega(J) := \nabla_J \bar{H}[\Psi_e](J) \;. \label{eq:defOmega}
\EE
To analyze equation (\ref{eq:LAVEtt}), we expand
\BE
 \MODE(J,\Theta,\tau) = \sum_{k\in\Z^2} \MODE_k(J,\tau)\,e^{ik\cdot\Theta}.
\EE
To obtain an equation for the Fourier coefficients, we define 
\BE
 K(J,J',\Theta) := \bar{G}\left(D(\beta_xJ_x,\beta_xJ'_x,\Theta_x),
                                D(\beta_yJ_y,\beta_yJ'_y,\Theta_y)\right)
\EE
with Fourier coefficients given by
\BE
 K_k(J,J') = \frac{1}{(2\pi)^2} \int_{\T^2} K(J,J',\Theta) e^{-ik\cdot\Theta} d\Theta.
\EE
Since $K$ is real and even in $\Theta_x$ and $\Theta_y$, $K_k$ is real,
 and since $K_k$ is real, $K_k=K_{-k}$. 
Also $K(J,J',\Theta)=K(J',J,\Theta)$, and 
\BE
 \bar{G}(D_x,D_y) = \sum_{k\in\Z^2} K_k(J,J')\,e^{ik\cdot(\Theta-\Theta')}.
\label{eq:GbarFC}
\EE
An easy calculation using (\ref{eq:GbarFC}) and  (\ref{eq:AVHAM}) gives
\BE
 \bar{H}[\MODE](J,\Theta) = (2\pi)^2\sum_{k\in\Z^2} e^{ik\cdot\Theta}\int_{\R_+^2} K_k(J,J')\MODE_k(J')\,dJ'.
\EE
Thus the Fourier modes determined by (\ref{eq:Mode}) are uncoupled and are given by
\BEA
 -i\PD{\tau}\MODE_k + k\cdot\Omega(J)\MODE_k && \nonumber \\
  \mp k\cdot \nabla f_e(J)\int_{\R_+^2} K_k(J,J')\MODE_k(J')\,dJ' &=& 0,
 \label{eq:LAVEtk} % LAVE in tau and k
\EEA
 where $f_e(J):=(2\pi)^2\Psi_e(J)$ is the equilibrium action density ($\int_{\R_+^2}f_e(J)dJ=1$).

There are two standard approaches to analyzing (\ref{eq:LAVEtk}): the Laplace transform approach and 
 the eigenvalue approach.

Taking the Laplace transform of (\ref{eq:LAVEtk})
 we obtain
\BEA
 (\omega-k\cdot\Omega(J))\;\LMODE_k(J) && \nonumber \\
 \pm\, k\cdot \nabla f_e(J) \int_{\R_+^2} K_k(J,J')\LMODE_k(J') \,dJ' & = & -if^\pm_k(J,0),
 \label{eq:LLAVEok} % Laplace LAVE in omega and k
\EEA
for $\Im \omega$ sufficiently large.
Here we use $-i\omega$ instead of the usual Laplace variable $s$,
 $\hat{f}_k(J,\omega)$ denotes the 
 Laplace transform of $f_k$, and $\Omega(J):= \bar{H}(\Psi_e)(J)'$.
In the eigenvalue approach, we
 look for solutions of the form
\BE
 \MODE_k(J,\tau) = \LMODE_k(J)\;e^{-i\omega\tau},
\EE
 which gives
\BEA
 (\omega-k\cdot\Omega(J))\;\LMODE_k(J) && \nonumber \\
 \pm\, k\cdot \nabla f_e(J) \int_{\R_+^2} K_k(J,J')\LMODE_k(J') \,dJ' & = & 0.
 \label{eq:LAVEok} % LAVE in omega and k
\EEA
The Laplace transformed equation (\ref{eq:LLAVEok}) has a non-homogeneous term,
 however, the left hand sides of equations (\ref{eq:LLAVEok}) and (\ref{eq:LAVEok}) are identical.
These equations are of the form 
\BE
a(x)\phi(x)-\lambda\int K(x,y)\phi(y)dy=f(x)
\label{eq:TKIE}
\EE
 and are called Fredholm 
 integral equations of the third kind (see p.2 \cite{Hochstadt}).
If $a$ is bounded away from zero it can be transformed into a Fredholm integral equation of 
 the second kind. 
Thus the primary interest in this third kind equation is in the case where $a$ has zeros and this
 is our case. 
The case where $a$ has zeros is complicated by the fact that there are generalized solutions 
 which are difficult to represent numerically.

Equations (\ref{eq:LLAVEok}) and (\ref{eq:LAVEok}) have analogues in both plasma physics and 
 other beam dynamics contexts.
For example, Crawford and Hislop \cite{CrHi} discuss the standard plasma problem in the periodic case,
 the case of this paper, 
 summarizing both the Landau and the van Kampen-Case solutions (\cite{Landau}, \cite{Kampen}, \cite{Case}).
Jackson \cite{Jackson} gives a nice presentation of the Landau approach in the non-periodic case.
The third kind integral equations are given in equations (17) and (23) of \cite{CrHi} and in equation (3.5)
 of \cite{Jackson}.
As is well known, the plasma problem leads to Landau damping and growth for certain equilibria
 depending on the size of the average density.
The stability analysis is facilitated by the dispersion function which uncouples the calculation of the poles
 of the solution from the calculation of the density itself. 

Two standard beam dynamics problems concern the longitudinal dynamics with wake fields for a coasting beam 
 and for a bunched beam.
The coasting beam case is completely analogous to the periodic plasma problem including a dispersion function
 and possible Landau damping for small beam current and an instability threshold at some critical current 
 after which there is Landau growth.
A recent discussion of the coasting beam problem in the context of coherent synchrotron radiation is
 given in \cite{VWRE}.
The third kind integral equation is given in equation (27) of that paper.
The emphasis in \cite{VWRE} is on the threshold for instability which occurs when a zero of the 
dispersion function reaches the real axis as the current increases from small values.
Landau damping is not discussed as it is not important for the stability discussion and furthermore would
 require an analytic continuation into the lower half $\omega$ plane which would require assumptions on
 the equilibrium (see p.7 in \cite{VWRE}).

The bunched beam case is more complicated as the Fourier modes do not decouple.
Furthermore, it appears at first sight that the 
 calculation of the instability threshold must be done in combination with the calculation 
 of the density. 
However, Warnock has introduced a regularization transformation which eliminates the continuous spectrum.
The resulting equation is then discretized leading to a determinant condition, independent of the density,
 which is analogous to the dispersion relation.
A convergence theorem would then make this rigorous.
This is discussed in \cite{WVE02}, where equation (11) is very similar to our equation (\ref{eq:LLAVEok}).
However the kernel of the integral equation is much different and, in fact, one expects a stability threshold.
It is in this paper that Warnock introduces his regularization transformation which eliminates the continuous
 spectrum and thus eliminates the numerical problem of representing generalized functions numerically.
More recent progress on the regularization is given in \cite{WSVE}.

Our equations (\ref{eq:LLAVEok}) and (\ref{eq:LAVEok}) are simpler than the longitudinal bunched beam equations
 in that the Fourier modes are uncoupled.
Also, our case is rather special in that it does not depend on the beam current as mentioned above.
In fact the $\pi$ and $\sigma$ eigen-modes are neutrally stable if $k\cdot\nabla f_e(J)\ne0$.
In this case, the transformation
\BE
   \LMODE_k(J) = \left| k\cdot\nabla f_e(J) \right|^{\frac{1}{2}} \;\SMODE_k(J)
\EE 
 leads to
\BEA
 (\omega-k\cdot\Omega(J))\,\SMODE_k(J) && \nonumber \\
 + \SETA \int_{\R_+^2}  \hat{K}_k(J,J')\SMODE_k(J')\,dJ' & = & 0,
 \label{eq:YIE} 
\EEA
 where $\hat{K}_k(J,J'):=|k\cdot\nabla f_e(J)|^{1/2}K_k(J,J')|k\cdot\nabla f_e(J')|^{1/2}$ and
  $\SETA:=\pm\,{\rm sgn}(k\cdot\nabla f_e)$.
Since the kernel $\hat{K}_k$ is real and symmetric, an eigenvalue $\omega$ must be real.
It is in fact a remarkable feature of the linearized AVE (\ref{eq:YIE}) for
 the case of $\nu=\nu^*$, $(\nu_x,\nu_y)$ far-from-low-order-resonance and for
 equilibria with $k\cdot\nabla f_e(J)\ne0$, that despite the presence of an
 amplitude dependent tune shift and a collective force, the modes show
 neither damping nor growth but instead are stable.
In \cite{WMPT,VESW01} we have given numerical evidence that the modes
 are indeed remarkably stable even in the fully nonlinear regime of tracking with
 equation (\ref{eq:KR}).

We have tried to show that eigensolutions $(\omega,\LMODE_k)$ of (\ref{eq:LAVEok}) for equilibria,
 which do not satisfy the above condition, $k\cdot\nabla f_e(J)\ne0$ (e.g.\ densities with two humps), must
 have $\omega\in\R$, but have been unsuccessful.
This leaves open the possibility of complex eigenvalues.
Since these must come in complex conjugate pairs, 
 there is the possibility of linearly (and thus nonlinearly) unstable solutions.

We have assumed that the two beams have the same nonresonant tunes and this is probably the reason that the 
 eigensolutions are neutrally stable.
Previous work indicates that when the tunes are different (a so-called tune split) or near-to-low-order 
 resonance there can be Landau damping or growth.
In \cite{ZY}, the authors develop a perturbation procedure in the near resonance case and argue that 
 there are regions of stability and instability (see Figure 2 of \cite{ZY} ).
Landau damping is also discussed in \cite{AL00}.
In \cite{WMPT}, we have seen evidence for Landau damping in the 2D phase space case (See Figures 5-7 of\cite{WMPT} ). 
A future problem for us is to determine the explicit form of the averaged equations in this case.

Equation (\ref{eq:YIE}) and the associated Laplace transformed equation can be rewritten as
 $(T-\omega I)\SMODE=h$.
Since $T$ is symmetric, we are looking for conditions which ensure that it is 
 a bounded selfadjoint operator on an appropriate Hilbert space.
Such operators have a well developed spectral theory.
For example, the spectrum is a compact subset of the real line contained
 in the interval $[m,M]$ where both $m$ and $M$ are finite spectral values and all spectral
 values are either in the point spectrum (eigenvalues) or 
 the continuous spectrum, thus the residual spectrum is empty \cite{KREY}.
Numerical results, in the section to follow, indicate that for $k\TRA=(1,0)$ or $(0,1)$ the 
 spectrum is the interval $[0,k\cdot\Omega(0)]$ for the 
 $\sigma$ mode with $0$ an eigenvalue and $[0,k\cdot\Omega(0)]\cup\{\omega_\pi\}$ 
 for the $\pi$ mode, with $\omega_\pi>k\cdot\Omega(0)$ an eigenvalue.
A possible explanation for the stability of the modes using the
 notion of the ``Landau resonance'' is that the modes can not resonantly
 exchange energy with a macroscopic fraction of the particles in the beam.
The $\sigma$ mode tune lies at the edge of the incoherent (continuous)
 spectrum towards infinite orbital amplitudes.
Any sensible phase space density falls off rapidly at large amplitudes
 (or even has compact support) so that the fraction of particles with tunes
 in resonance with the $\sigma$ mode tune vanishes.
The $\pi$ mode tune, in the studied parameter regime, lies considerably outside
 the incoherent spectrum and is thus even less able to dissipate its energy
 among the single particle trajectories or to draw energy from them.

\section{Numerical Results for {\large $\pi$} and {\large $\sigma$} Modes }\label{sc:num}

The analysis of the spectrum for equation (\ref{eq:YIE}) gives important information
about the $\pi$ and $\sigma$ modes.
In this section, we discuss properties of solutions of (\ref{eq:YIE})
and give our results on the numerical solution of this eigenvalue problem.

If $\omega$ is outside of the range of $k\cdot\Omega(J)$, 
  (\ref{eq:YIE}) can be reduced to an integral equation 
  of the second kind by a simple algebraic transformation. 
Conversely, if $\omega$ is in the range of $k\cdot\Omega(J)$, 
 then (\ref{eq:YIE}) it must be treated as a third kind equation.
Such equations have not been studied as extensively as Fredholm integral equations
 of the first and second kind.
A review of work up to 1973 and new results are given in \cite{Bart1} and more recent results are 
 contained in \cite{Bart2}. 
Most recently, we have become aware of \cite{Shulaia1,Shulaia2,Pereverzev}. 
However, to our knowledge, the case when $J$ is 2-dimensional 
 has not been discussed nor have convergent numerical schemes been developed.
As mentioned in the previous section the plasma problem and the coasting beam problems are of this type
 and have been studied extensively, however these are particularly simple.

We now discuss the commonly used, straightforward discretization for integral equations of the 
third kind as applied to our special case and give our numerical results.
At the end of this section we will discuss progress on work toward a convergent scheme.
In the straightforward approach, $J$ is put on a mesh and the integral is approximated by a
 simple quadrature method.
This leads to a finite dimensional matrix eigenproblem and seems to lead to reasonable results.
This approach has been used in 1D in the beam-beam interaction in \cite{YO00,AL00}, in the longitudinal
 bunched beam case in \cite{OY} and by us in the 1D beam-beam interaction, \cite{VESW01}. 
In \cite{VESW01} we obtained excellent agreement between the FFT spectra of the dipole modes
 in full blown simulations and the eigenvalues of a one degree-of-freedom version of (\ref{eq:YIE}).

We consider the special case of axially symmetric Gaussian beams, where
\BE
 f_e(J) = \frac{1}{\varepsilon^2} e^{-\,\frac{J_x+J_y}{\varepsilon}} \label{eq:ASGauss}
\EE 
 with $\pi\varepsilon$ being the rms emittance,
 and horizontal dipole modes where $k=(1,0)\TRA$.
With these choices equation (\ref{eq:YIE}) takes the explicit form
\BEA
 (\omega-\Omega_x(J))\,\SMODE_{1,0}(J) && \nonumber \\
 \mp\varepsilon^{-3}
 \int_{\R_+^2} e^{-\frac{J_x+J_y}{2\varepsilon}} K_{1,0}(J,J') e^{-\frac{J'_x+J'_y}
 {2\varepsilon}} \SMODE_{1,0}(J')\,dJ' & = &0.
 \label{eq:LAVEnum}
\EEA
We transform the actions $I_x=J_x/(1+J_x)$ and $I_y=J_y/(1+J_y)$, thereby mapping
 $\R^+\ra[0,1)$, and use a 60$\times$60 mesh. 
Even though the linearized averaged Vlasov equation (\ref{eq:LAVEnum})
 reduces the number of independent variables from four as in (\ref{eq:AVE}) to two,
 the evaluation of the functions  $\Omega(J)$ and $\hat{K}_k(J,J')$
 is in fact computationally expensive. 
The computation of $\Omega=\nabla\bar{H}[\Psi_e]$ involves a 6-fold
 integral at each point of the 2D mesh in $J$ and $\hat{K}_k$ involves 
 a 4-fold integral at each point of a 4D mesh in $(J,J')$.
Although we found a way to slightly simplify the calculation for general 
 $\Psi_e$, and reduce the 6-fold integral to a 5-fold integral, going to 
 larger meshes is quite expensive.
\begin{figure}
 \begin{center}
  \includegraphics[angle=0,width=8.5cm]{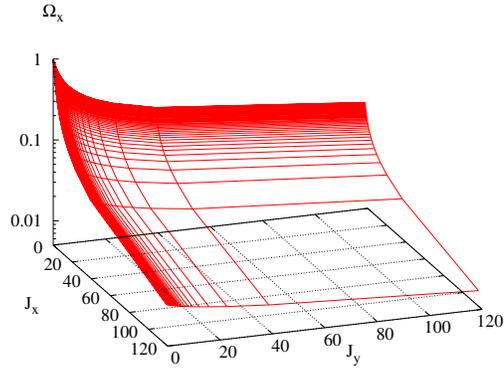}
 \end{center}
 \caption{$\Omega_x(J_x,J_y)$ for the equilibrium density (\ref{eq:ASGauss}).}
\label{fig:Omega}
\end{figure}
However, for the important particular choice of (\ref{eq:ASGauss})
  we found a very simple formula involving modified Bessel functions: 
\BEA
  \bar{H} [\Psi_e](J) = 
  \int_0^\infty\frac{dq}{2\varepsilon+q} && \nonumber \\
  \times\left(1- \exp\left(-\frac{J_x+J_y}{2\varepsilon+q}\right)
   I_0\left(\frac{J_x}{2\varepsilon+q}\right)
   I_0\left(\frac{J_y}{2\varepsilon+q}\right) \right). &&
\EEA
This formula has been known in the context of the weak-strong approximation of the 
 beam-beam tune shift, \cite{Sen}.
It is  straightforward to prove that
 $\lim_{J\ra 0} \Omega_x(J)=\lim_{J\ra 0} \Omega_y(J) = 1$, 
 and that the ranges of  $\Omega_x(J),\Omega_y(J)$ are both the interval $(0,1)$.
The latter is also the range of the continuous spectrum of (\ref{eq:YIE}).

\begin{figure}
 \begin{center}
  \includegraphics[angle=0,width=7.5cm]{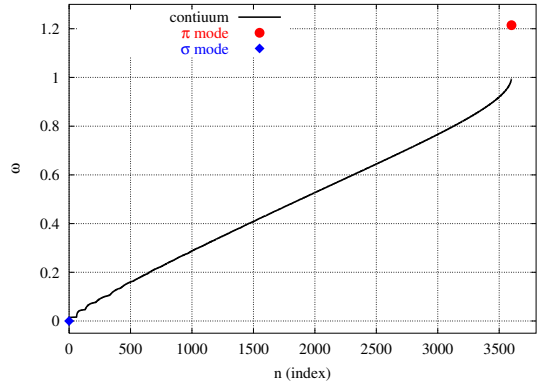}
 \end{center}
 \caption{The dipole spectrum of (\ref{eq:LAVEnum}).
  The black curve is the continuum, common to both modes,
  the red circle is the discrete eigenvalue for the $\pi$ mode and
  the blue diamond is the discrete eigenvalue for the $\sigma$ mode.}
 \label{fig:SigPi}
\end{figure}

$\Omega_x(J_x,J_y)$ is shown in Figure \ref{fig:Omega},
 and the spectrum of the finite dimensional approximation of (\ref{eq:LAVEnum}) 
 is shown in Figure \ref{fig:SigPi}.
The plot suggests that (\ref{eq:LAVEnum}) has a continuous spectrum,
 common to both the $\sigma$ and $\pi$ modes,
 which coincides with the range of $\Omega_x(J)$.
In addition, the $\sigma$ mode has a discrete eigenvalue $\omega=0$,
 and the $\pi$ mode has a discrete eigenvalue at $\omega\approx1.21$.
\begin{figure}
 \begin{center}
  \includegraphics[angle=0,width=8.5cm]{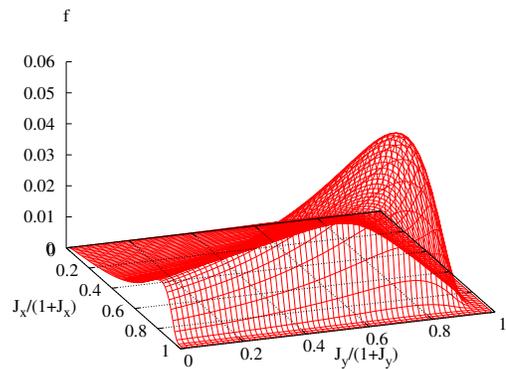}
 \end{center}
 \caption{The $\sigma$ mode eigenfunction of (\ref{eq:LAVEnum}).}
\label{fig:sigFunc}
\end{figure}
\begin{figure}
 \begin{center}
  \includegraphics[angle=0,width=8.5cm]{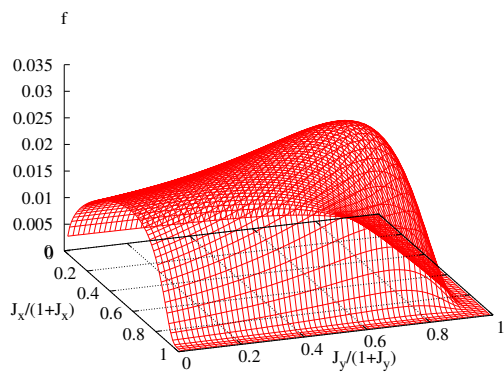}
 \end{center}
 \caption{The $\pi$ mode eigenfunction  of (\ref{eq:LAVEnum}).}
 \label{fig:piFunc}
\end{figure}
Figures \ref{fig:sigFunc} and \ref{fig:piFunc} show that these eigenvalues
 corresponds to regular eigenfunctions.
FFT spectra obtained by tracking, with our newly written
 parallel Perron-Frobenius code \cite{SobolThesis}, which tracks the phase space densities directly in
 4D phase space, and with our 4D weighted macro particle tracking code \cite{VESW01},
 shows pronounced peaks at tunes that correspond to these discrete eigenvalues.
This indicates an excellent agreement between these three completely different approaches.
We consider this a major feat.

\begin{figure}
 \begin{center}
  \includegraphics[angle=0,width=8.5cm]{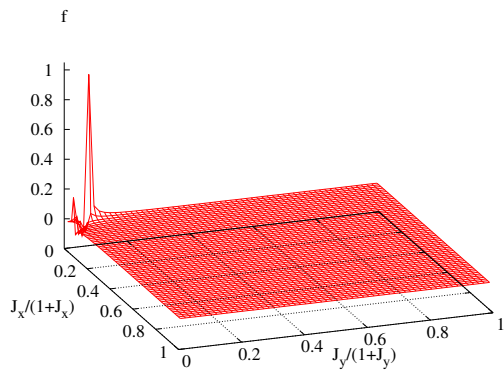}
 \end{center}
 \caption{An eigenfunction corresponding to an eigenvalue from the
          continuous spectrum of (\ref{eq:LAVEnum}).}
\label{fig:BadFn}
\end{figure}

Even though our results are quite satisfactory, we are interested in a convergent procedure.
A well established theory (See e.g., \cite{Kendall}) guarantees that
the above described finite dimensional approximation of the integral equation
converges as long as the operator is compact. 
However, as we mentioned before,
 Figure \ref{fig:SigPi} indicates the presence of a continuous spectrum, which
 a compact operator can not have.
In addition, the numerically computed ``eigenfunctions''  associated with the continuous
 spectrum show a singular behavior, as illustrated in Figure \ref{fig:BadFn}.
This is to be expected \cite{Bart1,Bart2} and thus these third kind integral equations are not
 well suited for numerical analysis in their standard form.
Therefore further numerical and analytical
analysis requires a special treatment of equation (\ref{eq:YIE}).
Such difficulties are common for equation (\ref{eq:TKIE}) 
 where $a(x)$ vanishes at least once inside of its domain.
The recent work for the longitudinal bunched beam Vlasov equation in \cite{WVE02, WSVE} mentioned above,
  suggests that the continuous spectrum can be eliminated by the Warnock regularization transformation
  $g(J,\omega) = (\omega-k\cdot\Omega(J)) \hat{g}(J,\omega)$.
This gives the nonlinear eigenvalue problem in $\omega$:
\BE
    g(J,\omega) - \eta \int_{\R_+^2} K_k(J,J')\frac{\hat{g}(J',\omega)}
     {\omega-k\cdot\Omega(J')} \,dJ'=0,
    \label{eq:NLEVP}
\EE
  where the solutions $(g(J,\omega),\omega)$ for $\eta=\pm1$ give the
  $\sigma$ and $\pi$ eigenmodes respectively.

As mentioned before, equation (\ref{eq:TKIE}) appears in other physics applications,
 and analytical results have been published in
 \cite{Bart1,Bart2,Shulaia1,Shulaia2,Pereverzev}.
We have begun a study of this problem in our particular case.
Specifically, the existence and uniqueness of solutions of equation (\ref{eq:YIE}) in the one
 and two-dimensional cases has been addressed in \cite{Sobol}.
In \cite{Sobol}, we consider functions which are continuous except for
 pole-like singularities at $J$ such that  $(\omega-k\cdot\Omega(J))=0$,
 and interpret the integrals over the singularities in the principal value sense.
Under certain assumptions,
 we proved a version of the Fredholm alternative theorem: the
 equation has a unique
 solution for any right-hand side iff the homogeneous version of this equation
 has only the trivial solution.
In addition, this framework may provide a numerically convergent scheme for solving (\ref{eq:YIE}).
Alternatively, we are looking for conditions so that discretization of the Warnock transformed problem will 
 lead to a convergent method.  
The theory of \cite{Sobol} is a first step.

\section{Summary and Future Work}\label{sc:summary}

An important aspect of collective beam-beam theory has been the study of the
 so-called $\pi$ and $\sigma$ modes.
The pioneering works of \cite{MS,CR00,YO00,AL00} represent a major advance;
 however approximations are made, the validity of which we would like to understand better.
For example, in \cite{YO00,AL00}, the starting point is a Vlasov equation with a delta function kick
 and the beam-beam kick is distributed around the ring by 
 smoothing the delta function.
The Vlasov equation is linearized about a function which is only an 
 approximate equilibrium and action-angle variables are introduced.
The Fourier modes in angle are not uncoupled at this stage but a horizontal dipole mode
 proportional to $\exp(i\Theta_x)$ is assumed.
This leads to an inconsistent equation.
To obtain a consistent equation for this mode an average over the angle
 variables is taken which, after a Fourier transform in time, leads to an integral 
 equation, the analog of equation (\ref{eq:YIE}).
In contrast, we start with the kick-lattice model to properly handle the
 delta function kick. 
Then we make only $\it one$ approximation, the averaging approximation, and in addition,
 as stated earlier, we believe we can give an upper bound on the error of approximation.
Our AVE has $\it{exact}$ equilibria and thus our exact problem has 
 quasi-equilibria in good agreement with our simulations.
The linearization about these equilibria leads to an equation, which in contrast to the above, 
 has uncoupled Fourier modes.
The Fourier modes satisfy an integral equation that is easily transformed to a formally 
 selfadjoint problem.
We are looking for conditions such that the associated operator is bounded and selfadjoint, a case which has a 
 well developed theory.
The standard computation of the $\pi$ and $\sigma$ dipole mode frequencies discussed in Section 6 is 
 in good agreement with density tracking based on equation (\ref{eq:EVS}), using both the PF and the 
 weighted macro-particle tracking methods.
However the standard numerical approach to numerically solving (\ref{eq:YIE}) does not converge 
 as the mesh size decreases beyond some limit and we are searching for convergent algorithms 
 such as that suggested in \cite{WVE02,Sobol}.

In summary,
 we have introduced a new model, the averaged Vlasov equation (\ref{eq:AVE}),
 for the collective beam-beam interaction in two degrees of freedom, which
 we believe has significant 
 potential for deepening our understanding of this important collective 
 effect.

Equation (\ref{eq:AVE}) was derived in the spirit of the rigorous
 analysis in \cite{DEV02}. We believe similar error bounds can be derived,
 thus we believe (\ref{eq:AVE}) gives a good approximation to the 
 basic dynamics of (\ref{eq:KR}).
In fact, we have checked the one degree-of-freedom analog of (\ref{eq:AVE}) with 
 two aspects of a full-blown density tracking approach,
 the existence of quasi-equilibria and the calculation
 of the  $\pi$ and $\sigma$ mode eigenvalues, with excellent results.
More importantly, we have checked the two degree-of-freedom AVE with two full-blown simulation codes
 and have also found excellent agreement in the calculation of $\pi$ and $\sigma$ mode eigenvalues.
Thus we have confidence in the model.

We have demonstrated its usefulness as a tool for calculating  $\pi$ and $\sigma$ mode eigenvalues
 and for clarifying the existence of quasi-equilibria.
In the case of leptons, progress has been made on the question of the existence of an equilibrium
 for the exact model \cite{WaEl}.
However,
it seems likely that exact equilibria do not exist in the hadron case as the underlying dynamics is 
 likely to be chaotic.
In addition, the AVE may lead to a faster algorithm for calculating the density evolution.
This is because the beam-beam parameters, $\zeta$ and $\zeta^*$, are small and thus 
 the time step in numerical integration of (\ref{eq:AVE}) can be 
 $O(1/ \rm{max}(\zeta,\zeta^*))$, which is significantly larger than one turn.
We propose to investigate this potential speed up by developing
 a numerical procedure to integrate (\ref{eq:AVE}).
As another example, the AVE will be useful in taking the next step beyond the linear theory to investigate 
 coupling between the $\pi$ and $\sigma$ Fourier modes.
In the Laplace-picture, we may be able to use the fixed point iteration scheme discussed
 for the plasma problem in \cite{HoIc} or in the eigen-picture presented here we may be able to use the 
 van Kampen-Case approach \cite{Kampen, Case}.
In the latter case, the work of \cite{CrHi} may be useful. 
Finally, we can investigate several other effects such as those discussed by
 Alexahin \cite{AL00}.
Some topics we are considering for future work are: (i) a study of the near-to-low-order resonance case 
 as we do in \cite{DEV02} (ii) adding another degree of freedom to study the effect of synchrotron motion 
 into the dynamics of equation (2) and (iii) a study of the effect of a tune split by letting $\nu^*_x=\nu_x
 +\zeta a$ and $\nu^*_y=\nu_y +\zeta b$ and then applying our averaging formalism.
Here $(a,b)$ allow us to vary the tune split in units of the kick parameter.
As in \cite{DEV02} we expect bifurcations as $a$ and $b$ vary.
Items (i) and (iii) likely includes the possibility of both Landau damping and growth.

Our main point is that we now have a model in which many 
 important collective beam-beam interaction effects can be studied in a more systematic way 
 then was previously available.

\section*{Acknowledgements}
Discussions with Bob Warnock, Yuri Alexahin, Scott Dumas and Tanaji Sen are 
 gratefully acknowledged. \\ \noindent
The work of JAE and AVS was supported by DOE contract DE-FG02-99ER41104.

\appendix

\section{ Derivation of the Beam-Beam Parameter}\label{sc:xi}

To calculate the kick we consider three inertial reference frames: 
 the rest frame of the synchronous particle of the unstarred bunch $F$,
 the lab frame $F'$, and the rest frame of the synchronous particle of the
 starred bunch $F^*$.
The coordinate axes of the three frames are parallel and oriented so that 
 viewed from the lab frame $F'$, $F$ is moving in the positive $z$ 
 direction with velocity $\beta c$ and $F^*$ is moving in the negative 
 $z$ direction with velocity $-\beta^* c$. 
In what follows we will consider the relative velocities of the particles with
 respect to the synchronous particle of their bunch as non--relativistic.

We consider the momentum change of an unstarred particle moving through the starred bunch.
In the kick approximation, we assume that the transverse spatial coordinates
 are not changed during the interaction, i.e.\ $r(t)=(x,y,z+ut)\TRA$.
Thus the change in transverse momentum of an unstarred particle passing through
 the starred beam measured in $F^*$ is
\BE
 \Delta P_\perp^*(x,y)=q\int_\R \BF{E}_\perp^*(x,y,z+ut)\,dt
                      =\frac{q}{u}\int_\R\BF{E}_\perp^*(x,y,z)\,dz,
\EE
 where $u$ is the speed of the unstarred particle in the starred frame,
\BE
 u = c\;\frac{\beta+\beta^*}{1+\beta\beta^*},
\EE
 and $\BF{E}_\perp^*$ is the transverse electric field of the starred bunch in 
 $F^*$.
Note that $\beta$ is basically the speed of an unstarred particle in the lab frame and
 that  $\BF{B}^*$ is approximately zero in $F^*$.

Since the 3-momentum $P^*$ is part of a 4-vector and the boosts involved are all in the
 longitudinal direction
\BE
 \Delta P_\perp = \Delta P'_\perp = \Delta P^*_\perp.
\EE 

From $\BF{E}_\perp^*=-\nabla_\perp\phi^*$ we have
\BEA
 \Delta P_\perp^*(x,y) & = & -\,\frac{N^*qq^*}{u 4\pi\epsilon_0}
  \int_\R dz \nabla_\perp \int_{\R^3}  \nonumber \\
 & & \times
  \rho_3^*(\tilde{r}_\perp,\tilde{z})
  \frac{d^2\tilde{r}_\perp d\tilde{z}}{\sqrt{\|r_\perp-\tilde{r}_\perp\|^2+(z-\tilde{z})^2}}
  \label{eq:DelPPerp} \\
 & = & -\,\frac{N^*qq^*}{u 4\pi\epsilon_0} 
  \int_{\R^3} \rho_3^*(\tilde{r}_\perp,\tilde{z})
  \int_\R dz \nabla_\perp \nonumber \\
 & & \times
  \frac{d^2\tilde{r}_\perp d\tilde{z}}{\sqrt{\|r_\perp-\tilde{r}_\perp\|^2+(z-\tilde{z})^2}}.
  \label{eq:XchLimits}
\EEA
One easily shows by direct integration that
\BE
 \int_\R dz \nabla_\perp 
 \frac{1}{\sqrt{\|r_\perp-\tilde{r}_\perp\|^2+(z-\tilde{z})^2}}
 = -\nabla_\perp \ln\|r_\perp-\tilde{r}_\perp\|^2, 
\EE
 which is independent of $z$.
Thus 
\BE
 \Delta P_\perp^*(x,y)  =  \frac{2N^*qq^*}{u 4\pi\epsilon_0}
 \nabla_\perp \underbrace{
    \int_{\R^2}\rho_2^*(\tilde{r}_\perp)\ln\|r_\perp-\tilde{r}_\perp\|}_{
    :=\POT[\rho_2^*](r_\perp)},
\EE
 where $\rho_2^*=\int_\R\rho_3^*\,dz$.
The interchange of limits going from (\ref{eq:DelPPerp}) to (\ref{eq:XchLimits})
 is justified for $\rho_3^*$ decaying sufficiently fast at $\infty$.
(The singularities in the integral are integrable.)

Since $\Delta P_\perp'=\Delta P_\perp^*$ the kick in the lab frame is
\BE
 \BM{c}\Delta p_x \\ \Delta p_y \EM = \frac{\Delta P_\perp^*}{p_0}.  
\EE
Since the speed of the unstarred synchronous particle in the lab frame is $\beta c$,
 we have $p_0=m\beta\gamma c$.
Thus the kick is
\BE
  \BM{c}\Delta p_x \\ \Delta p_y \EM = -\zeta \nabla_\perp \POT[\rho_2^*](r_\perp)
\EE
 where
\BE
 \zeta = \frac{2N^*qq^*}{4\pi m\epsilon_0 c^2} \,\frac{1}{\beta\gamma}\,
         \frac{1+\beta\beta^*}{\beta+\beta^*},
\EE
 which is the justification for equation (\ref{eq:KR}).

%\section*{References}

\end{document}